\documentclass[sigconf]{acmart}

\AtBeginDocument{%
  }
\setcopyright{acmlicensed}
\copyrightyear{2026}
\acmYear{2026}
\setcopyright{cc}
\setcctype{by}
\acmConference[MM '26]{Proceedings of the 34th ACM International Conference on Multimedia}{November 10--14, 2026}{Rio de Janeiro, Brazil}
\acmBooktitle{Proceedings of the 34th ACM International Conference on Multimedia (MM '26), November 10--14, 2026, Rio de Janeiro, Brazil}
\acmISBN{979-8-4007-2213-4/2026/11}
\acmDOI{10.1145/3767308.3836620}
\usepackage{tabularx}
\usepackage{array}
\usepackage{xurl}
\usepackage{balance}
\newcolumntype{Y}{>{\raggedright\arraybackslash}X}
\newcommand{\dataset}{AdoDAS}

\AtBeginDocument{}

\begin{document}

\title{AdoDAS: A Privacy-Preserving Multimodal Challenge for Adolescent Depression, Anxiety, and Stress Assessment}

\author{Zhaojie Luo}
\authornote{Both authors contributed equally to this research.}
\correspondingauthor
\orcid{0000-0002-4173-6319}
\email{luozhaojie@seu.edu.cn}
\affiliation{%
  \institution{Southeast University}
  \city{Nanjing}
  \state{Jiangsu}
  \country{China}}
\affiliation{%
  \institution{Shenzhen Loop Area Institute}
  \city{Shenzhen}
  \country{China}}

\author{Junkun Wang}
\authornotemark[1]
\orcid{0009-0004-9510-102X}
\email{k3nwong@seu.edu.cn}
\affiliation{%
  \institution{Southeast University}
  \city{Nanjing}
  \state{Jiangsu}
  \country{China}}

\author{Tianhua Qi}
\orcid{0009-0005-5780-9374}
\email{qitianhua@seu.edu.cn}
\affiliation{%
  \institution{Southeast University}
  \city{Nanjing}
  \state{Jiangsu}
  \country{China}}

\author{Yuxuan Wu}
\orcid{0009-0005-6481-1560}
\email{wyxuan@seu.edu.cn}
\affiliation{%
  \institution{Southeast University}
  \city{Nanjing}
  \state{Jiangsu}
  \country{China}}

\author{Xin Zhao}
\orcid{0009-0001-5221-9755}
\email{xinz@seu.edu.cn}
\affiliation{%
  \institution{Southeast University}
  \city{Nanjing}
  \state{Jiangsu}
  \country{China}}

\author{Tetsuya Takiguchi}
\orcid{0000-0001-5005-7679}
\email{takigu@kobe-u.ac.jp}
\affiliation{%
  \institution{Kobe University}
  \city{Kobe}
  \country{Japan}}

\author{Tomoko Matsui}
\orcid{0000-0003-3201-6106}
\email{tmatsui@ism.ac.jp}
\affiliation{%
  \institution{Shenzhen Loop Area Institute}
  \city{Shenzhen}
  \country{China}}

\author{Kun Qian}
\orcid{0000-0002-1918-6453}
\email{qian@bit.edu.cn}
\affiliation{%
  \institution{Beijing Institute of Technology}
  \city{Beijing}
  \country{China}}

\author{Fei Wang}
\orcid{0000-0002-5982-2303}
\email{feiwangster@gmail.com}
\affiliation{%
  \institution{Nanjing Medical University}
  \city{Nanjing}
  \country{China}}

\author{Shuqiong Wu}
\orcid{0000-0003-1501-9719}
\email{wushuqiong2kyoto@gmail.com}
\affiliation{%
  \institution{ The University of Osaka}
  \city{Osaka}
  \country{Japan}}

\author{Zhengjun Yue}
\orcid{0000-0002-1101-549X}
\email{z.yue@tudelft.nl}
\affiliation{%
  \institution{Shenzhen Loop Area Institute}
  \city{Shenzhen}
  \country{China}}

\author{Hiroshi Ishiguro}
\orcid{0000-0002-0805-7648}
\email{ishiguro@sys.es.osaka-u.ac.jp}
\affiliation{%
  \institution{The University of Osaka}
  \city{Osaka}
  \country{Japan}}

  \author{Xinyuan Qian}
\orcid{0000-0002-9511-6713}
\email{qianxy@ustb.edu.cn}
\affiliation{%
  \institution{University of Science and Technology Beijing}
  \city{Beijing}
  \country{China}}

\author{Haizhou Li}
\orcid{0000-0001-9158-9401}
\email{haizhouli@cuhk.edu.cn}
\affiliation{%
  \institution{The Chinese University of Hong Kong (Shenzhen)}
  \city{Shenzhen}
  \country{China}}
\affiliation{%
\institution{Shenzhen Loop Area Institute}
\city{Shenzhen}
\country{China}}

\renewcommand{\shortauthors}{Zhaojie Luo et al.}

\begin{abstract}
Adolescent depression, anxiety, and stress (D/A/S) call for scalable tools that complement, rather than replace, professional evaluation.
Under a privacy-preserving policy, the AdoDAS Grand Challenge withholds minors' raw recordings and distributes anonymized audio-visual representations and ASR-derived text.
Its 6{,}000 participants provide 24{,}000 segments across one scripted-reading and three open-response sessions.
Two tracks assess multi-task binary D/A/S screening and ordinal prediction of 21 DASS-21 item responses.
From 191 registrations, the final leaderboards included 95 eligible screening teams and 64 item-prediction teams.
Audio-visual baselines achieved 0.4604 mean F1 and 0.2675 mean Quadratic Weighted Kappa; leading submissions reached 0.5921 and 0.2776.
Representative systems emphasize cross-session modelling, temporal multimodal fusion, psychometric structure, and task-aware calibration.
\end{abstract}



\maketitle

\section{Introduction}
Mental-health conditions affect a substantial share of adolescents, with depression and anxiety among the leading causes of illness and disability in this age group~\cite{merikangas2009epidemiology,kieling2024worldwide,who2025adolescent}.
Early assessment is nevertheless constrained by reliance on self-report instruments, limited specialist capacity, and the difficulty of observing symptoms consistently across settings.
The Depression Anxiety Stress Scales (DASS) distinguish three related negative emotional states, and DASS-21 provides 21 ordinal items organized into Depression, Anxiety, and Stress subscales~\cite{lovibond1995dass,norton2007dass21}.
Speech timing and prosody~\cite{10446191,li2025low,qu2023disentangling,wang2025enhancing,qu2025disentanglement,lu2026speaker}, facial and body dynamics~\cite{van2007body,de2010standing,chen2025disenemo,11610589}, and verbal behaviour~\cite{fussell2002verbal,qi25_interspeech,eggins2004introduction,qi2026affectspeech} may provide complementary evidence, but computational outputs in this setting are screening signals and research measurements, not diagnoses.

Existing corpora and shared tasks established interview-based depression analysis, standardized audio-visual evaluation, clinical audio/physiological resources, and in-the-wild video settings~\cite{gratch2014distress,ringeval2019avec,cai2022modma,yoon2022dvlog}.
Recent benchmarks further study personality-aware and longitudinal user-level multimodal depression detection~\cite{mpdd_challenge,mud3,liu2025multi,alghowinem2016multimodal,tao2024depmstat,zhang2025assessing}.
However, large-scale adolescent assessment creates a distinct governance problem: raw voices and faces can reveal biometric identity even after obvious metadata are removed~\cite{han2020voice,ravi22_interspeech,wang23pa_interspeech}.
Few benchmarks combine multiple elicitation sessions, both coarse and item-level DASS targets, and a release policy designed to reduce direct exposure of minors' recordings.
Cross-language and cross-cultural studies also show that depression models may not transfer uniformly across populations and languages~\cite{abdelkadir-etal-2024-diverse,you25_interspeech}.

Because Depression, Anxiety, and Stress are related but distinct dimensions, subscale screening and constituent-item prediction are not interchangeable.
Coarse screening captures broad D/A/S risk, whereas ordinal item prediction retains symptom-specific variation and response order; together they expose shared and finer-grained structure without implying diagnosis.

The four sessions provide complementary elicitation conditions: scripted reading constrains lexical content, whereas open responses elicit personal narratives and spontaneous verbal and non-verbal behaviour.
Participant-linked sessions support subject-level cross-session evaluation.

The representation-only release makes privacy--utility a benchmark question: which multimodal and psychometric signals remain useful when anonymized text and derived features replace minors' raw recordings?
It reduces direct biometric exposure while testing what task-relevant information survives the interface.

The \dataset{} Challenge addresses this gap with three contributions.
First, it provides a subject-disjoint benchmark of 6{,}000 participants using standardized derived representations and anonymized ASR text rather than raw recordings.
Second, it defines complementary hidden-test tracks for D/A/S screening and DASS-21 item prediction, together with feature-only audio, video, and audio-visual baselines.
Third, it summarizes participation, results, and representative high-performing methods to identify transferable lessons and remaining limitations. Resources are available through \href{https://drive.google.com/drive/folders/1tBPUQN6g2g3qvcwxjYVprlYdqoxH14T2?usp=drive_link}{\textbf{AdoDAS DATA}}\footnote{\url{https://drive.google.com/drive/folders/1tBPUQN6g2g3qvcwxjYVprlYdqoxH14T2?usp=drive_link}} and the \href{http://adodas.hai-lab.cn/}{\textbf{AdoDAS Project Page}}\footnote{\url{http://adodas.hai-lab.cn/}}.

\section{Challenge Design}
\subsection{Dataset and Protocol}
The dataset contains 6{,}000 child and adolescent participants and 24{,}000 audio-video segments collected through a controlled school-based protocol.
Each participant contributes four sessions: a standardized reading passage (A01) and three open-ended responses (B01--B03) concerning the previous day, a happy memory, and a sad memory.
The subject-disjoint train/validation/test split contains 4{,}200/600/1{,}200 participants and 16{,}800/2{,}400/4{,}800 segments, respectively.
The complete cohort spans 10 anonymized schools and 250 anonymized school--class groups.
All four sessions from a participant remain in the same split and share participant-level questionnaire labels.
This grouping prevents session-level leakage: a model cannot observe one session from a test participant during training and then exploit participant-specific traits in another.
The 70/10/20 participant split preserves all 10 schools in each partition, so the principal evaluation question is generalization to unseen participants under broadly matched collection conditions.

\begin{figure*}[t]
  \centering
  \includegraphics[width=1\textwidth]{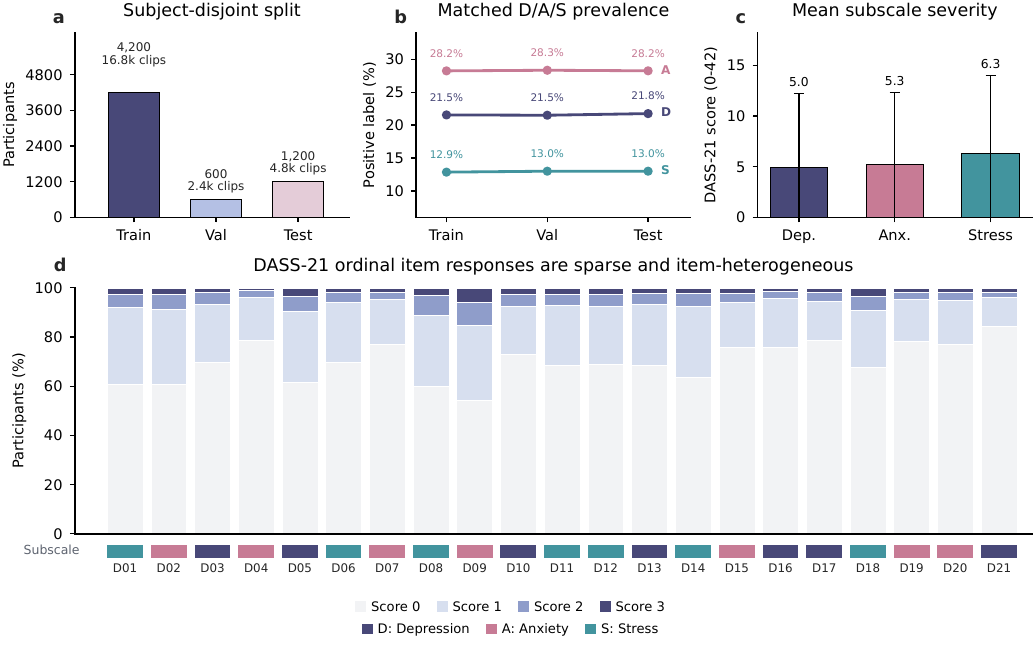}
  \caption{\dataset{} split sizes, D/A/S prevalence, DASS-21 subscale means (whiskers: one standard deviation), and item-response distributions.}
  \Description{Four panels show split sizes, D/A/S prevalence, DASS-21 subscale means, and item-response distributions.}
  \label{fig:dataset_overview}
\end{figure*}

Ground truth is derived from DASS-21.
For Track A1, each subscale is mapped to a binary target: \texttt{Normal} is negative and mild-or-above is positive.
Across the cohort, positive prevalence is 21.58\% for Depression, 28.25\% for Anxiety, and 12.90\% for Stress, with closely matched prevalence across the three splits.
For Track A2, the original item responses in $\{0,1,2,3\}$ are retained.
Their concentration at lower severity levels makes majority-oriented error measures insufficient by themselves and motivates an ordinal agreement metric.
Score zero is the majority response for every item, ranging from 54.40\% for d09 to 84.47\% for d21; non-zero responses are therefore highly item-dependent.
The reconstructed Depression, Anxiety, and Stress means are 4.95, 5.26, and 6.30 on the conventional doubled 0--42 DASS-21 scale.
These distributions make the two tracks complementary: A1 compresses each subscale into a coarse screening label, whereas A2 retains symptom-level variation and the ordered distance between response categories.
Figure~\ref{fig:dataset_overview} summarizes the split, prevalence, subscale, and item distributions.

\subsection{Privacy-Preserving Release}
Raw audio, video, and identifiable frames are not distributed.
The release provides anonymized identifiers, temporal metadata, frame-level or pooled audio-visual representations, and anonymized ASR transcripts for the three open-response sessions.
Audio representations include log-Mel/MFCC descriptors, voice-activity statistics, eGeMAPS functionals, and speech self-supervised embeddings; video representations include quality signals, head pose, facial behaviour, body pose, global motion, and visual self-supervised embeddings.
The sequence streams are aligned to a common 25~Hz timeline, while pooled descriptors and ASR text support low-cost and semantic modelling.
This interface shifts comparison away from raw-encoder scale toward temporal, multimodal, cross-session, and psychometric modelling.

Collection followed the applicable institutional and school procedures for research with minors; legal guardians provided written consent and students provided assent.
Withholding raw recordings reduces direct exposure but does not establish anonymity or formal privacy: ASR text and speech or facial representations may retain identity, institution, recording-condition, or demographic information~\cite{han2020voice,ravi22_interspeech,wang23pa_interspeech,chen2025faceprivacy}.
The interface also prevents end-to-end adaptation of raw-signal encoders and may remove fine-grained cues.
Results therefore measure learning under this release policy, not the performance ceiling of raw multimodal sensing or the absence of re-identification risk.

\subsection{Tracks and Evaluation}
Both tracks use the hidden test set and an organizer-controlled scorer.
Submissions are aligned by anonymized participant identifiers and must contain exactly the reference identifier set.
Track A1 requires probabilities for Depression, Anxiety, and Stress.
The scorer applies a fixed threshold of 0.5, calculates binary F1 for each target, and ranks teams by their mean; mean AUROC~\cite{lobo2008auc} is the tie-breaker.
The fixed threshold supports deterministic comparison but is not a clinically selected operating point.
Track A2 requires integer predictions from 0 to 3 for all 21 DASS-21 items.
Teams are ranked by mean item-level Quadratic Weighted Kappa (QWK)~\cite{vanbelle2016new}, with mean absolute error (MAE; lower is better)~\cite{willmott2005advantages} as the tie-breaker.
QWK rewards ordered agreement while accounting for chance and penalizes distant ordinal errors more strongly; MAE provides an interpretable check on absolute response distance, consistent with evidence that ordinal structure matters in speech-based depression modelling~\cite{zuo25_interspeech}.
Both tracks use macro-style averaging across their targets, preventing one subscale or frequently occurring item from defining the complete score.
Submission validation rejects duplicate or missing identifiers, non-finite probabilities, out-of-range A1 values, and non-integer A2 responses before scoring.

\section{Baseline and Challenge Protocol}
The official baseline operates on the released audio-visual features without a text branch.
Feature-group adapters project heterogeneous inputs into a common width; six-layer residual dilated temporal convolutions encode each modality; and attentive statistics pooling uses validity masks together with VAD and video-quality signals.
The resulting audio and video summaries are fused per session, and masked mean pooling followed by an MLP aggregates the four sessions into a participant representation.
These choices reflect established evidence for temporal/domain-adaptive speech models, acoustic--lexical fusion, and learned depression representations~\cite{huang20g_interspeech,villatorotello21_interspeech,wang22z_interspeech}.
A1 uses three sigmoid outputs, whereas A2 uses 21 CORAL-style ordinal heads with validation-selected decoding parameters.
The organizers fixed the split, input schema, submission validation, and scoring software across teams; the hidden reference labels remained inside the scoring environment.
The public implementation specifies feature selection, masks, optimization, random seed, validation selection, and A2 decoding, providing an executable reference for the protocol.

\begin{table}[t]
  \centering
  \caption{Official hidden-test baselines.}
  \label{tab:baseline}
  \small
  \begin{tabular}{llcc}
    \toprule
    Track & Input & Primary & Tie-break \\
    \midrule
    A1 & Audio & F1 .3959 & AUROC .6500 \\
    A1 & Video & F1 .4093 & AUROC .6749 \\
    A1 & Audio+Video & F1 .4604 & AUROC .7169 \\
    \midrule
    A2 & Audio & QWK .1260 & MAE .5675 \\
    A2 & Video & QWK .1758 & MAE .4700 \\
    A2 & Audio+Video & QWK .2675 & MAE .4679 \\
    \bottomrule
  \end{tabular}
\end{table}

Video exceeded audio on both primary metrics, and audio-visual fusion performed best, improving mean F1 by 0.0511 and mean QWK by 0.0917 over video.
Stress, the least prevalent A1 target, was weakest for the selected video baseline (F1 0.3067, versus 0.4390 for Depression and 0.4821 for Anxiety), showing how a mean can conceal target-level difficulty.
These compact reference systems verify the protocol and quantify modality gains on frozen organizer-provided representations; they do not measure raw-signal fine-tuning, whose performance can depend on task, language, transcript quality, and available evidence~\cite{gomezzaragoza25_interspeech,maji25_interspeech,deng25b_interspeech}.

\section{Participation and Results}
The challenge received 191 registration records, corresponding to 157 case-insensitive deduplicated team names and 97 normalized institutions.
After removing withdrawn entries, 95 teams were eligible for A1 and 64 for A2.
There were 105 unique leaderboard teams: 54 entered both tracks, 41 only A1, and 10 only A2.
Eligible A1 scores had a dense middle and a longer lower tail, with median F1 0.4236 and best F1 0.5921.
A2 had median QWK 0.2126 and a much more compressed high-performing group.
Among the 54 dual-track teams, the primary metrics had a Spearman correlation of 0.4429.
This moderate association suggests that useful representations and validation practices transfer between tasks, while leaving substantial room for track-specific modelling.

\begin{table}[t]
  \centering
  \caption{Organizer baseline and leading hidden-test results.}
  \label{tab:results}
  \scriptsize
  \setlength{\tabcolsep}{2.5pt}
  \begin{tabularx}{\columnwidth}{@{}Yrrrr@{}}
    \toprule
    System & F1 $\uparrow$ & AUROC $\uparrow$ & QWK $\uparrow$ & MAE $\downarrow$ \\
    \midrule
    Organizer A+V & .4604 & .7169 & .2675 & .4679 \\
    HITSZ\_HLT & \textbf{.5921} & .7652 & .2740 & \textbf{.4392} \\
    ustc-ac & .5853 & \textbf{.8019} & -- & -- \\
    cog-simulation & .4747 & .7051 & -- & -- \\
    HEALTH FIRST & .4710 & .7320 & \textbf{.2776} & .4807 \\
    Ritsumei & .4665 & .7282 & .2717 & .4695 \\
    Neuromorphic\_Computing & -- & -- & .2726 & .4767 \\
    daemon & -- & -- & .2660 & .4585 \\
    \bottomrule
  \end{tabularx}
\end{table}

Table~\ref{tab:results} combines the leaderboards without implying one aggregate ranking; a dash means that a system is outside that track's top five.
HITSZ-HLT led A1 with 0.5921 mean F1, a 28.6\% relative improvement over the fused baseline, while HEALTH FIRST led A2 with 0.2776 mean QWK, a 3.8\% relative improvement.
In A1, ranks one and two differed by 0.0067 F1 but rank two exceeded rank three by 0.1107; the entire A2 top-five range was only 0.0116 QWK.

\section{Top-Performing Methods and Lessons}
Table~\ref{tab:methods} contrasts three strong systems: HITSZ-HLT coupled cross-session and frame-level audio-visual interaction with staged correction; HEALTH FIRST modelled 21 ordinal items with query heads and validation-selected experts; and Ritsumei connected multimodal evidence, ASR-derived summaries, and confidence-aware item refinement.
Their use of long-range linguistic structure and multimodal fusion is consistent with prior interview-based modelling~\cite{xezonaki20_interspeech,villatorotello21_interspeech,burdisso23_interspeech}.

\begin{table}[t]
  \centering
  \caption{Representative high-performing systems.}
  \label{tab:methods}
  \scriptsize
  \begin{tabularx}{\columnwidth}{lY}
    \toprule
    Team & System characteristics \\
    \midrule
    HITSZ\_HLT & Cross-session Transformer; frame-level A--V attention; session-aware ASR-text adaptation; A1-guided item correction; staged task-head adjustment. \\
    HEALTH FIRST & Dual Conformers; cross-session fusion; query-aligned 21-item CORAL heads; validation-selected item experts. \\
    Ritsumei & DynaBridge multimodal/semantic summaries; item-to-D/A/S reconstruction; confidence-aware item refinement. \\
    \bottomrule
  \end{tabularx}
\end{table}

Four lessons recur across the strongest systems.
First, \emph{cross-session modelling} is useful because scripted reading offers standardized acoustic and visual conditions while open responses provide more variable affective and semantic evidence.
Second, \emph{temporal multimodal fusion} is consistently supported by both the baseline comparison and top systems; missingness-aware aggregation is particularly important for heterogeneous feature streams.
Third, \emph{psychometric structure} connects the tracks: item predictions can reconstruct D/A/S subscale evidence, while coarse risk estimates can regularize item decisions.
Fourth, \emph{calibration and post-processing} are part of the measured system, especially for A2 under QWK and imbalanced ordinal labels.

\section{Discussion and Conclusion}
\dataset{} establishes a participant-disjoint benchmark of 6{,}000 adolescents, four sessions, and complementary screening and item-level tasks without distributing raw recordings.
Audio-visual baselines outperformed unimodal systems, and 105 teams produced eligible results.
The best A1 system raised mean F1 from 0.4604 to 0.5921, whereas the best A2 system raised mean QWK from 0.2675 to 0.2776.
A1 formed a distinct top-two tier, while the A2 top five spanned only 0.0116 QWK.
The moderate cross-track correlation ($\rho=0.4429$) and recurring cross-session, multimodal, DASS-aware, and calibration mechanisms suggest partially shared capability with task-specific bottlenecks, but heterogeneous submissions without common ablations do not support causal attribution.

These results remain screening research signals, not diagnoses.
Withholding raw recordings reduces direct exposure but is not a formal privacy guarantee because text and learned representations can retain identity or context.
The split tests unseen participants under matched collection conditions, not unseen schools, cultures, devices, or clinical populations; cross-corpus studies likewise report degradation under dataset and demographic shift~\cite{botelho22_interspeech,dumpala25_interspeech}.
Future work should therefore prioritize consent-compatible site/device holdouts, target- and item-level calibration, A1--A2 psychometric consistency, and identity-leakage tests.
Overall, \dataset{} provides a bounded, evidence-based starting point for privacy-preserving multimodal adolescent D/A/S research.

\clearpage
\begin{acks}
This work was supported in part by the National Natural Science Foundation of China under Grant No. 62576096, in part by the Fundamental Research Funds for the Central Universities under Grant No. 2242026K30053, and in part by the Young Scientists Fund of State Key Laboratory of Digital Medical Engineering under Grant No. 202501.

Besides, we sincerely thank all the participants, the data collection team, and the platform support team.
\end{acks}

\bibliographystyle{unsrtnat}
\setlength{\vfuzz}{1.5pt}
\balance
\bibliography{references}

\end{document}